\documentstyle[12pt]{article}
\begin{document}
\baselineskip 24pt
\hfuzz=1pt
\setlength{\textheight}{8.5in}
\setlength{\topmargin}{0in}
\begin{centering}
\LARGE {\bf A simple quantum 
oblivious transfer protocol}
\\  \vspace{.75in}
\large {M. Ardehali}\footnote[1]
{email: ardehali@mel.cl.nec.co.jp}\,$^,$\footnote[2]
{Research Laboratories,
NEC Corporation,
Sagamihara, 
Kanagawa 229
Japan} 
\\ \vspace{.75in}
\end{centering}

\begin{abstract}
A simple and efficient
protocol for quantum oblivious transfer is proposed. The
protocol can easily be implemented with present 
technology  and 
is secure against cheaters with unlimited computing
power provided the receiver does not have the technology to
store the particles for an 
arbitrarily long period of time. The proposed protocol is a 
significant improvement over the previous protocols.
Unlike the protocol of Cr\'epeau and Kilian which is secure if
only if the
spin of the particle is measured along the $x$ or the $y$ axis,  
the present protocol is
perfectly secure no matter along which axes the spin of the particles 
are measured, and
unlike the protocol of Bennett {\em et al.} which requires tens of
thousand of particles, the present protocol requires only two
particles.
\vspace{.5in}
\end{abstract}
\pagebreak

In 1970, Wiesner \cite{1} wrote a highly innovative paper about quantum
cryptography \cite{2}, introducing a new branch of Physics
and computation. In his original paper, he also
introduced the concept of 
{\em Multiplexing}, which was later rediscovered by Rabin \cite{3},
and is now
usually called {\em Oblivious Transfer} (OT) \cite{4}.
The concept of OT
has turned out to be a very useful tool in designing cryptographic
protocols, and has been used for quite a while as a standard primitive
tool for constructing more complex protocols.
Let us briefly describe the OT protocol:
\\
1 - Alice knows one bit $\lambda$, where $\lambda$ is either $1$ or
$-1$ \cite{5}.
\\
2 - Bob obtains bit $\lambda$ from Alice with probability 0.5.
\\
3 - Bob knows whether or not he obtained bit $\lambda$.
\\
4 - Alice does not learn whether or not Bob obtained bit $\lambda$.

Cr\'epeau and Kilian \cite {6} have suggested that OT
can simply be achieved by having
Alice send Bob a single spin
$\frac{\displaystyle1}{\displaystyle2}$ particle,
for example an electron, encoding the 
OT bit into the spin of the particle
along the horizontal
or vertical axis (or encoding the OT bit
in polarization of the photon along the horizontal
or diagonal axis). Bob then 
randomly chooses the $x$ or the $y$ axis,
and measures the spin of the particle along that axis.
Finally Alice tells Bob the correct axis. 
This simple protocol is
secure if and only if Bob measures the spin of the particle along 
the horizontal or the vertical axis. For example if 
Bob measures the spin of the particle along the diagonal axis,
he will then 
obtain a considerable amount
of partial information about Alice's bit \cite {6}.
Bennett {\em et al.} \cite{7}
have proposed a protocol for quantum OT
which is free from this disadvantage;
but their protocol is rather inefficient, requiring tens of thousands
of particles to be sent and received for a simple decision making.
It is worth noting that all previous quantum oblivious protocols are 
insecure against EPR attack.

In this paper, we propose a simple and  efficient protocol for
quantum OT. 
The protocol is a considerable improvement over the
previous protocols. Unlike the protocol of 
Bennett {\em et al.}, the present protocol
requires only two particles to be
sent for a simple decision making, and unlike the protocol of
 Cr\'epeau and Kilian, the proposed protocol is perfectly secure
no matter along which axes the spin of the particles are 
measured.

The proposed protocol consists of the following steps:
\\
(1) Alice and Bob agree that $\lambda$ is encoded
in the product of the spin of the two particles along
the horizontal axis or along the vertical axis, i.e.,
$\lambda$ is encoded
in $b_1b_2$, where $b_1$ and $b_2$ are
spins of the first and the second particles
along the horizontal axis, or
$b_1$ and $b_2$ are
spins of the first and the second particles
along the vertical axis
(here horizontal axis refers to $x$ or $-x$ axes
and vertical axis refers to $y$ or $-y$ axes).
They also agree that $b_1=1$ ($b_2=1$)
indicates that spin of the first (second) particle is along 
$0$ or $\displaystyle \frac {\pi}{2}$ axis, and  
$b_1=-1$ ($b_2=-1$)
indicates that spin of the first (second) particle is along 
$\pi$ or $\displaystyle \frac {3\pi}{2}$ axis. For example if 
$\lambda=1$, and if Alice decides to encode $\lambda$ in the
product of the spin of the two particles along the 
horizontal axis, then she prepares two particles with their
spins along the $x$ axis, or two particles with their
spins along the $-x$ axis.
Similarly if 
$\lambda=-1$, and if Alice decides to encode $\lambda$ in the
product of the spin of the two particles along the 
vertical axis, then she prepares two particles with 
spin of the first particle
along the $y$ axis and spin of the second particle along the
$-y$ axis or spin of the the first particle
along the $-y$ axis and spin of the second particle along the
$y$ axis. 
\\
($2$) Alice encodes
$\lambda$ in
$b_1b_2$
and sends Bob the two particles.
\\
($3$) Bob measures the spin of both particles randomly along
the $x$ axis or
along the $y$ axis.
\\
($4$) Alice asks Bob if his measurements have been successful. If Bob
says no, then Alice goes to step 2. If Bob says yes, then Alice 
tells him only one of the following two alternatives:
\\
($i$) $\lambda$ is encoded in
the product of the spin of the two particles along the horizontal axis,
\\
($ii$) $\lambda$ is encoded in the
product of the spin of the two particles along the vertical axis,

We now show that if Bob does not have the technology
to store the particles
until step $4$, then the above
oblivious transfer protocol is secure
against cheaters with unlimited computing power.
First note that if Bob is honest, 
then the oblivious transfer
protocol can succeed without any difficulty. 
For example, assume $\lambda$ is
encoded in the 
product of the spin of the two particles along the
horizontal axis. If Bob measures the 
spins of the particles along the $x$ axis, then he learns the 
value of $\lambda$; but if he measures the spins of the particles
along the $y$ axis, then 
he does not gain
any information about $\lambda$.

Now consider a cheating Bob who measures the spin of
the first particle along
axis $\vec{a}$ at angle $\alpha$ with respect to the $x$ axis,
and measures the
spin of the second particle
along axis $\vec{b}$ at angle $\beta$ with respect to the $x$ axis.
The result of his measurement on the first photon
can be represented by 
a random variable $b'_1$ which takes values in the set $\{1,-1\}$. 
According to the standard rules of quantum theory
\begin{eqnarray} 
p\,\left(b'_1=b_1\right)= \cos^{2}\left(\frac{\alpha-\theta}{2}
\right), \qquad
p\,\left(b'_1=-b_1\right)= \sin^{2}
\left(\frac{\alpha-\theta}{2}\right),
\end{eqnarray}
where $\theta$ 
is the angle at which the spin of the first particle is measured by 
Alice and is in the set
$\left\{0,\displaystyle \frac{\pi}{2},\pi,
\displaystyle\frac{3\pi}{2}\right\}$.
Similarly the result of his
measurement on the second photon can be represented
by a random variable $b'_2$ taking values in the set $\{1,-1\}$. Again
according to quantum theory
\begin{eqnarray} \nonumber
&&p\,\left(b'_2=1 \mid b_2=1\right)= \cos^{2}\left(\frac{\beta-\phi}{2}
\right), \qquad
p\,\left(b'_2=-1 \mid b_2=-1\right)= \cos^{2}\left(\frac{\beta-\phi}{2} 
\right), \\ \nonumber
&&p\,\left(b'_2=1 \mid b_2=-1\right)= 
\sin^{2}\left(\frac{\beta-\phi}{2} \right),
\qquad
p\,\left(b'_2=-1 \mid b_2=1\right)= 
\sin^{2}\left(\frac{\beta-\phi}{2} \right), \\
\end{eqnarray}
where $\phi$ 
is the angle at which the spin of the first particle is measured by
by Alice and is in the set
$\left\{0,\displaystyle \frac{\pi}{2},\pi,
\displaystyle\frac{3\pi}{2}\right\}$.

Without loss of generality, first
we assume $\lambda=1$. We now
asks the following
question: Given that $\lambda=1$, what is the probability that
Bob will obtain
$\lambda'=1$, i.e., what is $p\left (\lambda'=1 \mid \lambda=1 \right)$?
To answere this question, we note that $\lambda$ is encoded along the
horizontal and vertical axes with equal probability,  i.e., 
\begin{eqnarray}  \nonumber
p\,\left(\lambda'=1 \mid \lambda=1 \right) &= &
p\,\left(b'_1b'_2=1 \mid b_1b_2=1 \right)  \\ \nonumber
&=&p\,(H) \, p\,\left(b'_1b'_2=1 \mid b_1b_2=1, H \right) \\ 
&+&p\,(V)\, p\,\left(b'_1b'_2=1 \mid b_1b_2=1, V \right),
\end{eqnarray} 
where $H$ means $\lambda$ is encoded along the horizontal axis
and $V$ means $\lambda$ is encoded along the vertical axis. 
We now note that
\begin{eqnarray} \nonumber
p\,\left(b'_{1}b'_{2}=1 \mid b_{1}b_{2}=1, H \right) &= &
p\,\left( b_{1}=1,b_{2}=1 \right) 
p\,\left(b'_{1}b'_{2}=1 \mid b_{1}=1,b_{2}=1, H \right) \\ \nonumber
&+&p\,\left( b_{1}=-1,b_{2}=-1 \right) 
p\,\left(b'_{1}b'_{2}=1 \mid b_{1}=-1,b_{2}=-1, H \right), \\ \nonumber
p\,\left(b'_{1}b'_{2}=1 \mid b_{1}b_{2}=1, V \right) &= &
p\,\left( b_{1}=1,b_{2}=1 \right) 
p\,\left(b'_{1}b'_{2}=1 \mid b_{1}=1,b_{2}=1, V \right) \\ \nonumber
&+&p\,\left( b_{1}=-1,b_{2}=-1 \right) 
p\,\left(b'_{1}b'_{2}=1 \mid b_{1}=-1,b_{2}=-1, V \right). \\
\end{eqnarray} 
Since Alice encodes $\lambda$ along $x$ and $-x$ with equal
probability, 
\begin{eqnarray} 
p\,\left( b_{1}=-1,b_{2}=-1 \right) =\frac{1}{2}, \qquad
p\,\left( b_{1}=1,b_{2}=1 \right) =\frac{1}{2},
\end{eqnarray} \nonumber
we thus have
\begin{eqnarray} \nonumber
&&p\,\left(b'_{1}b'_{2}=1 \mid b_{1}b_{2}=1, H \right) = 
\frac{1}{2}  \, p\,\left(b'_1=1 \mid b_1=1, H \right)
p\,\left(b'_2=1 \mid b_2=1, H \right) \\ \nonumber
&&+\frac{1}{2} \, p\,\left(b'_1=1 \mid b_1=-1, H \right)
p\,\left(b'_2=1 \mid b_2=-1, H \right)  \\ \nonumber
&&+ \frac{1}{2} \, p\,\left(b'_1=-1 \mid b_1=1, H \right)
p\,\left(b'_2=-1 \mid b_2=1, H \right) \\ \nonumber
&&+ \frac{1}{2} \, p\,\left(b'_1=-1 \mid b_1=-1, H  \right)
p\,\left(b'_2=-1 \mid b_2=-1, H \right) \\ \nonumber
&&= \frac{1}{2} \, {\Bigg [} \cos^{2}\left(\frac{\alpha}{2}\right)
\cos^{2}\left(\frac{\beta}{2}\right) 
+ \sin^{2}\left(\frac{\alpha}{2}\right)
\sin^{2}\left(\frac{\beta}{2}\right) \\ \nonumber
&&+ \sin^{2}\left(\frac{\alpha}{2}\right)
\sin^{2}\left(\frac{\beta}{2}\right)
+ \cos^{2}\left(\frac{\alpha}{2}\right)
\cos^{2}\left(\frac{\beta}{2}\right)  {\Bigg ]}. \\
\end{eqnarray} 
Similarly
\begin{eqnarray} \nonumber
&&p\,\left(b'_{1}b'_{2}=1  \mid b_{1}b_{2} = 1 , V \right) = 
\frac{1}{2} \, p\,\left(b'_1=1 \mid b_1=1, V \right)
p\,\left(b'_2=1 \mid b_2=1, V \right) \\ \nonumber
&&+\frac{1}{2} \, p\,\left(b'_1=1 \mid b_1=-1, V \right)
p\,\left(b'_2=1 \mid b_2=-1, V \right)  \\ \nonumber
&&+\frac{1}{2} \, p\,\left(b'_1=-1 \mid b_1=1, V \right)
p\,\left(b'_2=-1 \mid b_2=1, V \right) \\ \nonumber
&&+ \frac{1}{2} \, p\,\left(b'_1=-1 \mid b_1=-1, V \right)
p\,\left(b'_2=-1 \mid b_2=-1, V \right) \\ \nonumber
&&= \frac{1}{2} {\Bigg [}
\cos^{2}\left(\frac{\alpha}{2} - \frac{\pi}{4}\right)
\cos^{2}\left(\frac{\beta}{2}  - \frac{\pi}{4}\right)
+ \sin^{2}\left(\frac{\alpha}{2}  - \frac{\pi}{4}\right)
\sin^{2}\left(\frac{\beta}{2}  - \frac{\pi}{4}\right) \\ \nonumber
&&+ \sin^{2}\left(\frac{\alpha}{2}  - \frac{\pi}{4}\right)
\sin^{2}\left(\frac{\beta}{2}  - \frac{\pi}{4}\right) 
+\cos^{2}\left(\frac{\alpha}{2}  - \frac{\pi}{4}\right)
\cos^{2}\left(\frac{\beta}{2}  - \frac{\pi}{4}\right) {\Bigg ]}.
\\
\end{eqnarray}
Using  Eqs. (3), (6) and (7), and noting that 
$p \,(H)=p \, (V)=\frac{\displaystyle 1}{\displaystyle 2}$, we obtain
\begin{eqnarray}  \nonumber
&&p\,\left(\lambda'=1 \mid \lambda=1 \right) 
=\frac{1}{2} {\Bigg [} \cos^{2}\left(\frac{\alpha}{2}\right)
\cos^{2}\left(\frac{\beta}{2}\right) 
+\sin^{2}\left(\frac{\alpha}{2}\right)
\sin^{2}\left(\frac{\beta}{2}\right)\\ \nonumber
&&+\cos^{2}\left(\frac{\alpha}{2} - \frac{\pi}{4}\right)
\cos^{2}\left(\frac{\beta}{2}  - \frac{\pi}{4}\right) 
+\sin^{2}\left(\frac{\alpha}{2}  - \frac{\pi}{4}\right)
\sin^{2}\left(\frac{\beta}{2}  - \frac{\pi}{4}\right) {\Bigg ]}.
\end{eqnarray}
Using the following simple trigonometric relations,
\begin{eqnarray} \nonumber
&&\cos \left(\gamma_1-\gamma_2\right)=
\cos\gamma_1\, \cos\gamma_2+\sin\gamma_1\,\sin\gamma_{2}, \\
&&\sin\left(2\gamma\right)=
2 \sin\gamma cos\gamma, \qquad
\cos^2\gamma=\frac{1+\cos(2\gamma)}{2},
\end{eqnarray}
Eq. $(7)$ may be simplified to
\begin{eqnarray} \nonumber
p\,\left(\lambda' =1 \mid \lambda =1\right) &= &
\frac{1}{2} +\frac{1}{4} \left (\cos\, \alpha
\cos\,\beta  +
\sin\,\alpha
\sin\,\beta \right) \\ 
&=& 
\frac{1}{2} +\frac{1}{4} \cos\,\left(\alpha - \beta\right).
\end{eqnarray}
Obviously
\begin{eqnarray} \nonumber
p\,\left(\lambda' =-1 \mid \lambda =1\right) &=& 
1-p\,\left(\lambda' =1 \mid \lambda =1\right) \\
&=&\frac{1}{2} -\frac{1}{4} \cos\,\left(\alpha - \beta\right).
\end{eqnarray}
The probability that $\lambda'=1$ ($\lambda'=-1$) 
given that $\lambda=1$ is
maximized (minimized) if $\alpha = \beta$ and the maximum (minimum)
value is
$\displaystyle \frac{3}{4}
\left( \displaystyle \frac{1}{4} \right)$.
Thus the best strategy for a cheating Bob is to measure the spin of
both particles along the same axis, in which case he would obtain
as much information as an honest Bob who measures the spin of both
particles along the $x$ or along the $y$ axis, i.e., a cheating Bob
can not gain any more information about $\lambda$ than an honest Bob.

Next we assume that $\lambda=-1$. 
We now ask the following
question: Given that $\lambda=-1$, what is the probability that
$\lambda'=1$, i.e., 
what is $p\left (\lambda'=1 \mid \lambda=-1 \right)$?
To answere this question, again we note that
$\lambda$ is encoded along the
horizontal and vertical axes with equal probability,  i.e.,
\begin{eqnarray} \nonumber
p\,\left(\lambda'=1 \mid \lambda=-1 \right) &= &
p\,\left(b'_1b'_2=1 \mid b_1b_2=-1 \right)  \\ \nonumber
&=&p \,(H)\, p\,\left(b'_1b'_2=1 \mid b_1b_2=-1, H \right) \\ 
&+&p\,(V)\, p\,\left(b'_1b'_2=1 \mid b_1b_2=-1, V \right).
\end{eqnarray} 
We now note that
\begin{eqnarray} \nonumber
p\,\left(b'_{1}b'_{2}=1 \mid b_{1}b_{2}=-1, H \right) &= &
p\,\left( b_{1}=1,b_{2}=-1 \right)
p\,\left(b'_{1}b'_{2}=1 \mid b_{1}=1,b_{2}=-1, H \right) \\ \nonumber
&+&p\,\left( b_{1}=-1,b_{2}=1 \right)
p\,\left(b'_{1}b'_{2}=1 \mid b_{1}=-1,b_{2}=-1, H \right), \\ \nonumber
p\,\left(b'_{1}b'_{2}=1 \mid b_{1}b_{2}=-1, V \right) &= &
p\,\left( b_{1}=1,b_{2}=-1 \right)
p\,\left(b'_{1}b'_{2}=1 \mid b_{1}=1,b_{2}=-1, V \right) \\ \nonumber
&+&p\,\left( b_{1}=-1,b_{2}=1 \right)
p\,\left(b'_{1}b'_{2}=1 \mid b_{1}=-1,b_{2}=1, V \right). \\
\end{eqnarray} 
Since Alice encodes $\lambda$ along $x$ and $-x$ with equal
probability,
\begin{eqnarray}
p\,\left( b_{1}=1,b_{2}=-1 \right) =\frac{1}{2}, \qquad
p\,\left( b_{1}=-1,b_{2}=1 \right) =\frac{1}{2},
\end{eqnarray} 
we thus have
\begin{eqnarray} \nonumber
&&p\,\left(b'_{1}b'_{2}=1 \mid b_{1}b_{2}=-1, H \right) = 
\frac{1}{2} \, p\,\left(b'_1=1 \mid b_1=1, H \right)
p\,\left(b'_2=1 \mid b_2=-1, H \right) \\ \nonumber
&&+\frac{1}{2} \, p\,\left(b'_1=-1 \mid b_1=1, H \right)
p\,\left(b'_2=-1 \mid b_2=-1, H \right)  \\ \nonumber
&&+ \frac{1}{2} \, p\,\left(b'_1=1 \mid b_1=-1, H \right)
p\,\left(b'_2=1 \mid b_2=1, H \right) \\ \nonumber
&&+ \frac{1}{2}\,  p\,\left(b'_1=-1 \mid b_1=-1, H  \right)
p\,\left(b'_2=-1 \mid b_2=1, H \right) \\ \nonumber
&&= \frac{1}{2} {\Bigg [} \cos^{2}\left(\frac{\alpha}{2}\right)
\sin^{2}\left(\frac{\beta}{2}\right) +
\sin^{2}\left(\frac{\alpha}{2}\right)
\cos^{2}\left(\frac{\beta}{2}\right) \\ \nonumber
&&+ \sin^{2}\left(\frac{\alpha}{2}\right)
\cos^{2}\left(\frac{\beta}{2}\right)
+\cos^{2}\left(\frac{\alpha}{2}\right)
\sin^{2}\left(\frac{\beta}{2}\right){\Bigg ]}. \\
\end{eqnarray}
Similarly
\begin{eqnarray} \nonumber
&&p\,\left(b'_{1}b'_{2}=1 \mid b_{1}b_{2}=-1, V \right) = 
\frac{1}{2} \, p\,\left(b'_1=1 \mid b_1=1, V \right)
p\,\left(b'_2=1 \mid b_2=-1, V \right) \\ \nonumber
&&+ \frac{1}{2} \, p\,\left(b'_1=-1 \mid b_1=1, V \right)
p\,\left(b'_2=-1 \mid b_2=-1, V \right)  \\ \nonumber
&&+ \frac{1}{2} \, p\,\left(b'_1=1 \mid b_1=-1, V \right)
p\,\left(b'_2=1 \mid b_2=1, V \right) \\ \nonumber
&&+ \frac{1}{2} \, p\,\left(b'_1=-1 \mid b_1=-1, V \right)
p\,\left(b'_2=-1 \mid b_2=1, V \right) \\ \nonumber
&&= \frac{1}{2} {\Bigg [}
\cos^{2}\left(\frac{\alpha}{2} - \frac{\pi}{4}\right)
\sin^{2}\left(\frac{\beta}{2}  - \frac{\pi}{4}\right)+
\sin^{2}\left(\frac{\alpha}{2}  - \frac{\pi}{4}\right)
\cos^{2}\left(\frac{\beta}{2}  - \frac{\pi}{4}\right) \\ \nonumber
&&+ \sin^{2}\left(\frac{\alpha}{2}  - \frac{\pi}{4}\right)
\cos^{2}\left(\frac{\beta}{2}  - \frac{\pi}{4}\right)
+\cos^{2}\left(\frac{\alpha}{2}  - \frac{\pi}{4}\right)
\sin^{2}\left(\frac{\beta}{2}  - \frac{\pi}{4}\right) {\Bigg ]}. \\
\end{eqnarray}
Again using the fact that
$p \, (H)=p \, (V)= \displaystyle \frac{1}{2}$
 and referring to Eqs. (11), (14), and (15), we have
\begin{eqnarray}  \nonumber
&&p\,\left(\lambda'=-1 \mid \lambda=-1 \right)
= \frac{1}{2} {\Bigg [} \cos^{2}\left(\frac{\alpha}{2}\right)
\sin^{2}\left(\frac{\beta}{2}\right)
+\sin^{2}\left(\frac{\alpha}{2}\right)
\cos^{2}\left(\frac{\beta}{2}\right)\\ 
&&+\cos^{2}\left(\frac{\alpha}{2} - \frac{\pi}{4}\right)
\sin^{2}\left(\frac{\beta}{2}  - \frac{\pi}{4}\right)
+\sin^{2}\left(\frac{\alpha}{2}  - \frac{\pi}{4}\right)
\cos^{2}\left(\frac{\beta}{2}  - \frac{\pi}{4}\right) {\Bigg ]}
\end{eqnarray}
Simplifying the above equation, we obtain
\begin{eqnarray} \nonumber
p\,\left(\lambda'=1 \mid \lambda=-1\right)&=&
\frac{1}{2} -\frac{1}{4} \left (\cos\, \alpha
\cos\,\beta  -
\sin\,\alpha
\sin\,\beta \right) \\
&=&
\frac{1}{2} -\frac{1}{4} \cos\,\left(\alpha - \beta\right).
\end{eqnarray}
Obviously
\begin{eqnarray} \nonumber
p\,\left(\lambda' =-1 \mid \lambda =-1\right) &=&
1- p\,\left(\lambda'=1 \mid \lambda=-1\right)\\
&=&\frac{1}{2} +\frac{1}{4} \cos\,\left(\alpha - \beta\right).
\end{eqnarray}
The probability that $\lambda'=1$ ($\lambda'=-1$)
given that $\lambda=-1$ is minimized
(maximized) if $\alpha = \beta$ and the minimum (maximum) value is
$\displaystyle \frac{1}{4}
\left (\displaystyle \frac{3}{4} \right)$.
Thus the best strategy for a cheating Bob is to measure the spin of
both particles along the same axis, in which case he would obtain
as much information as an honest Bob who measures the spin of both
particles along the $x$ or along the $y$ axis.
Finally it should be noted that
Bob can cheat by storing the particles until step $4$ and then perform
his measurements. This sophisticated attack, which is in principle
possible, is completely
infeasible at present or in the foreseeable future.

To summarize, a cryptographic protocol for quantum OT is proposed.
The protocol is a significant improvement over the previous protocols.
Unlike the protocol of Kilian and Cr\'epeau which is secure if and only
if the spin of the particles are measured along the horizontal or 
vertical axis, the present protocol is
secure no matter along which axis the spin of the particles are 
measured, and
unlike the protocol of Bennett {\em et al.} which requires tens of 
thousand of photons, the present protocol requires only two photons.
However, similar to previous protocols, the present protocol is not
secure against EPR attack.
The advantage of the present protocol is
that it is extremely simple, highly economical and is secure against
cheater with technology that is available today or in foreseeable 
future.

\pagebreak

\end{document}